\documentstyle[12pt]{article}
\begin{document}

\small \hoffset = -1truecm \voffset = -2truecm
\title{\bf Quasinormal modes of stringy black holes}
\author{
    Xin-zhou Li\footnote {e-mail address: kychz@shtu.edu.cn}\hspace{0.6cm}  Jian-gang Hao\hspace{0.6cm} Dao-jun Liu\\\footnotesize \it
    Department of physics,Shanghai Normal University, Shanghai 200234 ,China
  \\\footnotesize \it
    East China Institute for Theoretical Physics, Shanghai 200237,China
     }
\date{}
\maketitle

\begin{abstract}
   String theory is a promising candidate for a fundamental quantum theory of all interactions
   including Einstein gravity. Some solutions in
string theory can be interpreted as black holes. Using the
semi-analytic method and WKB method,the quasinormal modes(QNMs)of
1+1 dimensional black hole in string theory are studied.The QNMs
of 1+3 dimensional black hole in string theory are also calculated
through numerical approach.The numerical investigation has shown
that the late time gravitational oscillation of the black hole
under an external perturbaton is dominated by certain QNMs.
\end{abstract}

\vspace{9cm} \hspace{0.8cm} PACS number(s): 04.20.Jb, 04.80.-y
\newpage

\noindent \textbf{1.\hspace{0.4cm}Introduction }\bigskip

The pioneering work on the stability of a black hole was carried
out by Regge and Wheeler[1], who studied the linear perturbation
of schwarzschild black hole. Further work on this problem led to
the study of quasinormal modes(QNMs)[2] and their role in the
response of black hole to external perturbation. There are
definite relations between the parameters of the black hole and
its QNMs. Since the gravitational radiation excited by the black
hole oscillation is dominated by its QNMs, one can determine the
parameters of a black hole by analyzing the QNMs in its
gravitational radiation. Therefore, besides their importance in
the analysis of the stability of the black holes, QNMs are
important in the search for black holes and their gravitational
radiation. The QNMs of black holes in the framework of general
relativity have been studied widely[3].

On the other hand, due to the discovery of the anomaly
cancellation for superstring[4], the construction of the heterotic
strings[5] and the seminal ideas on the compactification to four
dimensions[6], string theory has become the best candidate for a
fundamental quantum theory of all interactions including Einstein
gravity. The existence of a stable solution is important to the
general theory of the extended objects[7]. In string theory, some
solutions[8] can also be interpreted as black holes whose
parameters can be deduced from string theory by corresponding
compatifications. By studying the black hole in string theory,
people have successfully counted the black hole microstates, which
are the origin of black hole entropy. In this paper,we study the
QNMs of 1+1 dimensional black hole in string theory by the
semi-analytic method[9] and WKB method[10]. We also calculated the
QNMs of 1+3 dimensional stringy black hole through numerical
approach. Analyzing the QNMs of black hole in string theory
provides an alternative way to fix the parameters of the black
hole and thereby of string theory.

\bigskip

\noindent\textbf{ 2. 1+1 dimensional stringy black hole
solution}\bigskip

The exact solutions of string theory in the form of 1+1
dimensional black holes have attracted much interest[11]. These
stringy black holes can be obtained by non-minimally coupling the
dilaton to gravitational system. The action[12] is:

\begin{equation}
S=\int d^{2}x \sqrt{G}e^{-2\phi}\left[R-4(\nabla\phi^{})
^{2}-c-\frac{1}{4}F^{2}\right]
\end{equation}

\noindent where $F_{\mu \nu }$ is the Maxwell tensor, $\phi $ is
the dilaton field. By varying the action, one can obtain the
equations of motion:

\begin{equation}
R_{\mu \nu } = 2\phi _{;\mu ;\nu } + \frac{{1}}{{2}} F_{\mu \sigma
} F_{\nu }^{\sigma }\hspace{3cm}
\end{equation}

\begin{equation}
\left( {e^{ - 2\phi } F^{\mu \nu } } \right)_{;\nu } =
0\hspace{3cm}
\end{equation}

\begin{equation}
R + 4\phi _{;\rho } \phi ^{;\rho } - 4\phi _{;\rho }^{;\rho } - c
- \frac{{1}}{{4}}F^{2} = 0\hspace{3cm}
\end{equation}

\noindent Using schwarzschild-like gauge[11]

\begin{equation}
g_{\mu \nu }  = \left( {\matrix{{g\left( {r} \right)}&{0}\cr {0}&{
- \frac{{1}}{{g\left( {r} \right)}}}\cr }} \right)\hspace{3cm}
\end{equation}

\noindent one can solve the Eqs.(2-4) and the solutions are:

\begin{eqnarray}
\phi  &=& \phi _{0}  - \frac{{Q}}{{2}}r\hspace{3cm}\nonumber\\
 f &=& \sqrt {2} Qqe^{ - Qr}\left( { = F_{tr} } \right)\\
 g &=& 1 - 2me^{ - Qr} + q^{2} e^{ - 2Qr}\hspace{3cm}\nonumber
\end{eqnarray}

\noindent where $\phi_{0}$ and ${\it Q}$ are integration
constants. The asymptotic flatness condition requires $c=-Q^{2}$
and $Q>0$ . Parameters \textit{m} and \textit{q} can be considered
as the mass and charge of the black hole respectively. When
$m^{2}>q^{2}$ , the event horizon of the black hole lies at

\begin{equation}
 r_{\pm }  = \left( {\frac{{1}}{{Q}}} \right)ln\left( {m\pm \sqrt
{m^{2} - q^{2} } } \right)
\end{equation}

Now ,we consider the behaviors of the space time geometry under
the gravitational perturbation. We add the perturbing fields
$\varepsilon _{1}$ , $\varepsilon _{2}$ , $\varepsilon _{\phi } $
and $\varepsilon _{f} $ as follows:

\begin{equation}
ds^{2}  = \left( {g + \varepsilon _{1} } \right)dt^{2} - \left(
{\frac{{1}}{{g}} + \varepsilon _{2} } \right)dr^{2}
\end{equation}

\begin{equation}
 \phi \left( {r} \right) \to \phi \left( {r} \right) +
\varepsilon _{\phi } \left( {r,t} \right)
\end{equation}

\begin{equation}
 f\left( {r} \right) \to f\left( {r} \right) +
\varepsilon _{f} \left( {r,t} \right)
\end{equation}

\noindent We use the ans\"{a}tz:

\begin{equation}
\epsilon_{\phi}(r,t)=\epsilon_{\phi}(r)e^{i\omega t}
\end{equation}

\noindent Substituting Eqs.(8-10) into the equations of
motion(2-4), one can obtain the equations for the perturbing
fields

\begin{equation}
 g\phi ,_{r} \varepsilon _{2}  = 2\varepsilon_{\phi,r}e^{i\omega t}
  - \frac{{1}}{{g}}g,_{r} \varepsilon _{\phi }
\end{equation}

\begin{equation}
 \varepsilon _{f}  = \frac{{1}}{{2g}}f\varepsilon _{1}
+ \frac{{1}}{{2}}gf\varepsilon _{2}  + 2f\varepsilon _{\phi }
\end{equation}

\begin{equation}
 g^{2} \varepsilon _{\phi ,r,r}  = \omega ^{2}
\varepsilon _{\phi }
\end{equation}

\noindent Introducing the tortoise coordinate

\begin{equation}
 \frac{{dr_{\ast } }}{{dr}} = \frac{{1}}{{g\left( {r}
\right)}}
\end{equation}

\noindent We get the Regge-Wheeler equation for perturbation

\begin{equation}
\frac{d\psi(r_{*})}{dr_{*}}+[\omega^{2}-V(r_{*})]\psi(r_{*})=0
\end{equation}

\noindent where the effective potential is

\begin{equation}
 V\left( {r_{\ast } } \right) = \frac{{3}}{{4g\left(
{r} \right)^{2} }}\left[ {\frac{{dg\left( {r} \right)}}{{dr_{\ast
} }}} \right]^{2}  - \frac{{1}}{{2g\left( {r}
\right)}}\frac{{d^{2} g\left( {r} \right)}}{{dr_{\ast } ^{2} }}
\end{equation}

The complexity of the effective potential excludes exact analytic
solution for Eq.(23), one must therefore solve the equation by
certain approximation. Firstly, we take a method pioneered by
Mashhoon and co-workers[9], in which the effective potential is
replaced by a parametrized analytic potential whose simple exact
solutions are known . The parameters in the potential are obtained
by fitting the height , curvature and the asymptotic value of the
two potentials. The QNMs attained by this way agree with the
numerical results very well.

For effective potential Eq.(18) , we use Eckart potential to fit
it.

\begin{equation}
 V = V_{0} e^{2\mu } - V_{0} \left\{ {tanh\left[
 {\alpha \left( {r_{\ast 0 } - r_{\ast } } \right) + \mu }
\right] - tanh\mu } \right\}^{2} \times cosh^{2} \mu
\end{equation}

\noindent where $r_{\ast 0} $ is the value at which the potential
is maximum, the rest are free parameters which can be determined
by fitting Eckart potential and the effective potential.

Mashhoon and coworkers[12] has shown that the low-lying QNMs are
mainly determined by the maximum value of the effective potential
and its derivatives at maximum. After fitting, the three
parameters relate to the effective potential by

\[
V_{0}  =\frac{V_{max}}{\theta}
\]

\begin{equation}
\alpha=\frac{(1+\theta)V^{'''}_{max}}{6(\theta-1)V^{''}_{max}}
\end{equation}

\[
\mu=\frac{1}{2}\ln\theta
\]

\noindent where

\[
\theta=1+\frac{3V^{''}_{max}}{V^{'''}_{max}}\sqrt{\frac{-V^{''}_{max}}{2V_{max}}}
\]

\noindent $V_{max}$ , $V^{''}_{max}$ and $V^{'''}_{max}$ are the
maximum of the effective potential , its second and third
derivatives at maximum respectively.

By solving the Regge-Wheeler equation, in which the effective
potential has been replaced by Eckart potential, we can obtain the
QNMs

\begin{equation}
 \omega  = a\left( {1+ \delta } \right) + b\left( {1
- \delta } \right)i
\end{equation}

\noindent where

\[
a = \sqrt{{V_{0} cosh^{2} \mu  - \frac{{1}}{{4}}\alpha ^{2}
}},\hspace{1cm} b = \alpha \left( {n + \frac{{1}}{{2}}} \right)
\]

\begin{equation}
 \delta  =
\frac{{1}}{{2}}V_{0} \frac{{sinh\left( {2\mu } \right)}}{{a^{2} +
b^{2} }}
\end{equation}

\bigskip

To see wether this approach is reliable, we use another
approximate approach, in which we make use of the WKB method to
solve the equation and thereby obtain the QNMs. In WKB
approximation, we expend WKB functions to the third order and the
Taylor expansion near the turning points will be up to and
including terms in the sixth derivative of effective potential
$V(r_{*})$ . The QNMs is as follows

\begin{equation}
 \omega  = \sqrt {\left[ {V_{0}  + \left( { - 2V''_{0}
} \right)^{1/2} \tilde \Lambda } \right] - i\left( {n +
\frac{{1}}{{2}}} \right)\left( { - 2V''_{0} } \right)^{1/2} \left(
{1 + \tilde \Omega } \right)}
\end{equation}

\noindent where

\begin{equation}
 \tilde \Lambda \left( {n} \right) =
\frac{{1}}{{\left( { - 2V''_{0} } \right)^{1/2} }}\left[
{\frac{{1}}{{8}}\left( {\frac{{V_{0} ^{\left( {4} \right)}
}}{{V''_{0} }}} \right)\left( {\frac{{1}}{{4}} + \alpha ^{2} }
\right) - \frac{{1}}{{288}}\left( {\frac{{V'''_{0} }}{{V''_{0} }}}
\right)^{2} \left( {7 + 60\alpha ^{2} } \right)} \right]
\end{equation}

\begin{eqnarray}
\tilde \Omega \left( {n} \right) &=& \frac{1}{( { - 2V''_{0} }
)}\bigg[ {\frac{{5}}{{6912}}\bigg( {\frac{{V'''_{0} }}{{V''_{0}
}}} \bigg)}^{4}( {77 + 188\alpha ^{2} })- \frac{{1}}{{384}}\bigg(
{\frac{{V'''^{2} _{0} V_{0} ^{( {4} )} }}{{V_{0}^{'' 3} }}}
\bigg)( {51 + 100\alpha ^{2} } )
\hspace{3cm}\nonumber\\
& &+\frac{1}{2304}\bigg(\frac{V_{0}^{(4)}}{V_{0}^{''}}\bigg)^{2}
(67+68\alpha^{2})+\frac{1}{288}\bigg(\frac{V_{0}^{'''}V_{0}^{(5)}}{V_{0}^{''2}}\bigg)
(19+28\alpha^{2})-\frac{1}{288}
\bigg(\frac{V_{0}^{(6)}}{V_{0}^{''}}\bigg)(5+4\alpha^{2}) \bigg]
\end{eqnarray}

\begin{equation}
\alpha  = n + \frac{{1}}{{2}}
\end{equation}

\noindent where, $\it{V_{0}}$ is the maximum of the effective
potential $\it{V(\it{r_{*}})}$, $\it{V_{0} ^{\left( {n}\right)}}$
is the nth derivative of $\it{V(r_{*})}$ at its maximum value . In
table 1, we list the numerical results of the QNMs obtained
through the two approaches mentioned above and their agreements
are very good, especially for the imaginary parts.

Tab.1 The QNMs of 1+1 dimensional stringy black hole ($m=q=1$)

\hspace{2cm} \special{bmp:d:/Tab.1.bmp x=12cm y=12cm}

\bigskip
\noindent \textbf{3. QNMs of 1+3 dimensional black hole in string
theory}
\bigskip

In the framework of string theory, the 1+3 dimensional black hole
solution[13] is

\begin{equation}
 ds^{2}  =  - f^{ - 1/2} \left( {r} \right)\left( {1 -
\frac{{r_{0} }}{{r}}} \right)dt^{2}  + f^{1/2} \left( {r}
\right)\left[ {\left( {1 - \frac{{r_{0} }}{{r}}} \right)^{ - 1}
dr^{2}  + r^{2} \left( {d\theta ^{2}  + Sin^{2} \vartheta d\phi
^{2} } \right)} \right]
\end{equation}

\noindent where

\begin{equation}
 f\left( {r} \right) = \left( {1 + \frac{{r_{0}
Sinh^{2} \delta _{2} }}{{r}}} \right)\left( {1 + \frac{{r_{0}
Sinh^{2} \delta _{5} }}{{r}}} \right)\left( {1 + \frac{{r_{0}
Sinh^{2} \delta _{6} }}{{r}}} \right)\left( {1 + \frac{{r_{0}
Sinh^{2} \delta _{p} }}{{r}}} \right)
\end{equation}

\noindent $r_{0}$ is the horizon of the black hole . $\delta_{2}$
, $\delta_{5}$ , $ \delta_{6}$ and $ \delta_{p}$ are parameters
result from the compactification of higher dimensions.

\noindent By doing the transformation

\begin{equation}
 f^{1/2} \left( {r} \right)r^{2}  = \rho ^{2}
\end{equation}

\noindent we can transform the metric(26) into a familiar form

\begin{equation}
ds^{2}=B(\rho)dt^{2}-A(\rho)d\rho^{2}-\rho^{2}d\Omega
\end{equation}

\noindent where

\begin{equation}
 A\left( {\rho } \right) = \frac{{1}}{{r\left( {\rho }
\right)\left[ {r\left( {\rho } \right) - r_{0} } \right]}}\rho
^{2} \left( {\frac{{dr\left( {\rho } \right)}}{{d\rho }}}
\right)^{2}
\end{equation}

\begin{equation}
 B\left( {\rho } \right) = \frac{{1}}{{\rho ^{2}
}}r\left( {\rho } \right)\left[ {r\left( {\rho } \right) - r_{0} }
\right]
\end{equation}

The generic form of the Regge-Wheeler equation corresponding to
the external perturbation is
\begin{equation}
\frac{d\psi(r_{*})}{dr_{*}}+[\omega^{2}-V(r_{*})]\psi(r_{*})=0
\end{equation}

\noindent where

\begin{equation}
 V\left( {\rho } \right) = \left( {l - 1}
\right)\left( {l + 2} \right)\frac{{\Theta \Delta }}{{\rho ^{4} }}
- \frac{{\Delta }}{{\rho }}\frac{{d}}{{d\rho }}\left(
{\frac{{\Delta }}{{\rho ^{4} }}} \right)
\end{equation}

\begin{equation}
 \Theta  = \left(
{\frac{{dr\left( {\rho } \right)}}{{d\rho }}} \right)^{4}
\end{equation}

\begin{equation}
 \Delta  = \frac{{1}}{{\rho ^{2} }}r\left( {\rho } \right)\left[ {r\left(
{\rho } \right) - r_{0} } \right]\frac{{1}}{{\frac{{dr\left( {\rho
} \right)}}{{d\rho }}}}
\end{equation}

\noindent The generalized tortoise coordinate takes the generic
form

\begin{equation}
 \frac{{d}}{{dr_{\ast } }} = \frac{{\Delta }}{{\rho
^{2} }}\frac{{d}}{{d\rho }}
\end{equation}

\bigskip
We now restrict ourselves to the special case in which $\delta
_{2}  = \delta _{5}  = \delta _{6}  = \delta _{p}$ [14].

\noindent Let
\begin{equation}
Sinh^{2} \delta _{2}  = Sinh^{2} \delta _{5}  = Sinh^{2} \delta
_{6}  = Sinh^{2} \delta _{p}  = a^{2}\end{equation}

\noindent The effective potential becomes

\begin{eqnarray}
V({\rho }) &=& \frac{{3( {1 + 6a^{2}  + 6a^{4} })r_{0}^{2}
}}{{\rho ^{4} }} - \frac{{7a^{2}( {1 + 3a^{2}  + 2a^{4} }
)r_{0}^{3} }}{{\rho ^{5} }} + \frac{{4a^{4}( {1 + a^{2} } )^{2}
r_{0}^{4} }}{{\rho ^{6} }}\nonumber\\ & &- \frac{{5r_{0}( {1 +
2a^{2} })}}{{\rho ^{3} }} + \frac{{( {l^{2}  + l - 2} )\sqrt {(
{\rho  - a^{2} r_{0} } )( {r_{0} - \rho  + a^{2} } r_{0} }
)}}{{\rho ^{4} }} + \frac{{2}}{{\rho ^{2} }}
\end{eqnarray}

Then we can calculate the QNMs of 1+3 dimensional black hole , and
the numerical results ( Where we chose $r_{0}  = a = 1$) are
listed in Table 2.

Tab.2  The QNMs of 1+3 dimensional stringy black hole through
numerical approach.

\hspace{2cm} \special{bmp:d:/Tab.2.bmp x=12cm y=12cm}

\bigskip
In this letter, we calculate the QNMs of 1+1 dimensional stringy
black hole through the semi-analytic method as well as WKB method
and obtain the QNMs of 1+3 dimensional stringy black hole through
numerical approach. The results have shown that the late time
gravitational oscillation of these objects under an external
perturbation is dominated by certain QNMs. we have establish a
connection between the parameters of stringy solution and the
resonant oscillation modes which are in principle observable.
Therefore, this result will be helpful for determining the
parameters of the stringy black hole.

\bigskip
This work is supported by the National Natural Science Foundation
of China under Grant No. 19875016 ,  the Doctorial Fund from the
Education Ministry of China under Grant No. 1999025110 and the
Foundation for the Development of Science and Technology of
Shanghai under the Grant No. 00JC14057.

\end{document}